\documentclass[11pt]{JHEP3}
\title{Branes wrapping black holes as a purely gravitational dielectric effect}
\author{Diego Rodr\'{\i}guez-G\'omez\\ Departamento de F\'{\i}sica, Universidad de Oviedo\\
Avda.Calvo Sotelo 18, 33007 Oviedo, Spain\\
\email{Diego.Rodriguez.Gomez@cern.ch}}
\abstract{In this paper we give a microscopical description of certain configurations of branes wrapping black hole horizons in terms of dielectric gravitational waves. Interestingly, the configurations are stable only due to the gravitational background. Therefore, this constitutes a nice example of purely gravitational dielectric effect.}
\preprint{FFUOV-05/06}

\begin{document}

\section{Introduction}

Recently, an interesting proposal for the $AdS_2/CFT_1$ correspondence was
raised in \cite{GSY}. From the expression for the $\mathcal{N}=2$ 4-dimensional black hole
partition function in \cite{OSV}, it was observed that the most natural way of
describing the black hole is in terms of fixed magnetic charges and a certain
ensemble of weighted electric charges. This led to the study of a system of
$N$ D0 branes (electric charges) in the background attractor geometry produced
by magnetic D4 flux (\cite{GSSY}). Putting all the observations together, it
was conjectured in \cite{GSY} that the superconformal quantum mechanics of 
D0 branes is the dual of string theory in the attractor geometry,
which is $AdS_2\times S^2\times \mathcal{M}$, being the horizon of the black
hole the $S^2$ and $\mathcal{M}$ a certain compact manifold (which
  must be Calabi-Yau if we want $\mathcal{N}=2$ SUSY).

In this background, there are several stable supersymmetric brane configurations which are wrapping
non-contractible cycles of $S^2\times \mathcal{M}$. Among them, it is of particular interest that of a D2 brane
wrapping the $S^2$ with some DBI flux on it, carrying momentum along $\mathcal{M}$. As it was shown in \cite{SSTY},
these configurations preserve one half of the supersymmetries, and, since they
do not have net D2 charge, they contribute only to D0 charge. Indeed, in view of the
Myers effect (\cite{M}), one can understand microscopically these
configurations in terms of dielectric D0 branes. Such brane
configurations provide a natural understanding of the black hole entropy, since it can be
regarded as the degeneracy of the ground state of the configuration, which
is a Lowest Landau Level.

In a subsequent paper (\cite{DGMSWZ}) it was realized that similar
configurations of branes wrapping horizons with similar dispersion relations
occur in many other black hole backgrounds. In this reference, the
dynamics of branes wrapping black hole horizons in geometries of the form
$AdS_m\times S^n\times \mathcal{M}$ was studied from the point of view of
the $DBI + CS$ effective action. The branes are assumed to wrap the sphere and, since
in general they carry momentum in $\mathcal{M}$, they couple to the
background potential. As in the $AdS_2\times S^2$ case, these p-branes do not contribute to net p-brane charge,
but to D0 (or momentum) charge.

In this note we will concentrate on a class of
these configurations\footnote{For further details about the properties of
  these configurations, such as the preserved supersymmetries and other
  interesting issues, we refer to \cite{DGMSWZ}.}, namely
the ones static in $\mathcal{M}$ (as we will see, due to this ansatz, the
considered branes do not
couple to the background potential). This ansatz implies that
our branes lie in the lowest energy level, which is a Lowest Landau Level. In particular, we will work with the 11 dimensional $AdS_3\times
S^2\times \mathcal{M}$ background, which can be seen as the near horizon geometry of the
11d black string, which is carrying momentum along a certain direction
$y$. This background leads, upon reduction along $y$, to a type IIA geometry
whose near horizon is precisely $AdS_2\times S^2\times \mathcal{M}$. We will
also study the type IIB $AdS_3\times S^3\times\mathcal{M}$ background. 

In both
the M-theory and type IIB setups we will have stable expanded p-branes which do not
carry p-brane charge but momentum charge along a direction cointained in the $AdS$. In a sense, this is very reminiscent of the giant gravitons
(\cite{giants}). Giant gravitons are expanded stable branes wrapping a
contractible cycle and orbiting in the background. Since the coupling to the background flux compensates
exactly the tension of the brane, they behave as a massless particle. Given that these branes carry momentum charge dissolved in their
worldvolume, in the view of the Myers effect (\cite{M}), it is
natural to conjecture that there should exist a microscopical description of
giant gravitons in terms of pointlike gravitons (gravitational waves) expanded
due to dielectric effect. In very much of the same spirit of this, in this note we will see that we can give
a microscopical description of these configurations using the action for
coincident gravitons given in \cite{JL2}\footnote{A microscopical description
  for a certain class of giant gravitons has been also discused from a slightly different point of view also in \cite{SJ}.}. 

As we have so far mentioned, due to the fact that the branes we will consider
are static in $\mathcal{M}$, in both the 11d and IIB case the configuration is stable with
no help of the form potential. In other words, this provides an example of
purely gravitational dielectric effect (\cite{BGSW}).

Purely gravitational dielectric effect was expected in \cite{BGSW} for time
dependent configurations in non-trivial backgrounds\footnote{For some examples
  of purely gravitatorial dielectric effect see \cite{gravdielef}.}. In this reference it is
also pointed that this effect could be of importance in the context of
black holes. Here we have a nice example of this, since in the type IIB and M-theory
cases the configurations we are dealing with carry momentum in a certain
direction (\textit{i.e.} are not static) and are stable only due to the metric
background, with no help of the fluxes.

The type IIA case appears, at first sight, somehow different, since we will
have to consider not gravitons but static (in global coordinates) D0 branes. We will see that this can be
understood from its 11d origin. Due to the assumption that the
configuration sits in the origin\footnote{Strictly speaking, we only need the branes to be static in $\mathcal{M}$. In any case, we can always choose this point to be at the origin.} of $\mathcal{M}$, there is no coupling to the
RR 3-form potential. In turn, there is only the monopolar coupling of each D0
brane to the RR 1-form potential. We can gain some understanding by working in
the full 11d geometry prior to the near horizon limit, where we will see that there are stable configurations of
dielectric gravitational waves moving in $y$ expanded only due to gravitational
effects. Upon reduction, the GW become dielectric D0 branes, and consequently these
configurations will give rise to their IIA counterparts in terms of D0 branes. From this point of view, the particular IIA configurations we
will study are due to nothing but the purely gravitational dielectric effect
but seen with the Kaluza-Klein glasses.

Although we will consider the most supersymmetric backgrounds in each
situation\footnote{In general we could consider M-theory in a certain manifold
  $\mathcal{M}$, whose near horizon is $AdS_3\times S^2\times\mathcal{M}$. Upon
  reduction we would have a geometry whose near horizon would be $AdS_2\times
  S^2\times \mathcal{M}$. If $\mathcal{M}$ were a $CY_3$ we would have a
  $\mathcal{N}=2$ background, while if $\mathcal{M}=T^6$ we have the maximal
  supersymmetry.}, since for simplicity we will assume the configurations to
be in the origin of the extra manifold, the discussion could be lift to the
$\mathcal{N}=2$ case. In addition, the simplification of considering the
configurations at the origin of $\mathcal{M}$ makes pretty obvious that,
even in the most general case, the purely gravitational dielectric effect is
playing a key role.

\section{The action for coincident gravitons}

In \cite{JL1} the first step towards a microscopical description of coincident
gravitons was made. The final theory for 11 dimensional coincident gravitons
with dielectric couplings in their worldvolume was given in \cite{JL2}. This
action has been successfully used, providing a microscopical description of
giant gravitons in a series of papers (\cite{JLR1},\cite{JLR2},\cite{JLR3})
with a perfect agreement with their macroscopic counterparts in terms of usual branes. More recently, this action was shown to contain as a limit
the BMN matrix model, and used to derive a ``strongly coupled'' version of the
BMN matrix model by means of a 9-11 flip, which allowed to give a
microscopical description of the M5 vacuum of the 11d Pp-wave
(\cite{LR}).

The worldvolume theory associated to $N$ coincident gravitational waves in
 M-theory is a $U(N)$ gauge theory, in which the vector field is associated to
 M2-branes (wrapped on the direction of propagation of the waves) ending on
 them \cite{JL2}. This vector field gives the BI field living in a set of
 coincident D0-branes upon reduction along the direction of propagation of the
 waves. 
 
 In this paper we will use a truncated version of the action in \cite{JL2}
 in which the vector field is set to zero, given that it will not play any
 role in the backgrounds that we will be discussing. This action is given by 
 $S=S^{BI}+S^{CS}$, with
 
\begin{equation}
 \label{MBI}
 S^{BI}=-T_0\int dt\ {\rm STr} \{ k^{-1}\sqrt{-P[E_{00}+E_{0i}
 (Q^{-1}-\delta)^i_k E^{kj}E_{j0}]{\rm det Q}}\}\ ,
 \end{equation}

\noindent where

 \begin{eqnarray}
 \label{mred}
 &&E_{\mu\nu}={\mathcal G}_{\mu\nu}+k^{-1}(i_k C^{(3)})_{\mu\nu}\ , \qquad
 {\mathcal G}_{\mu\nu}=g_{\mu\nu}-\frac{k_\mu k_\nu}{k^2}\ , \\
 &&Q^i_j=\delta^i_j+ik[X^i,X^k]E_{kj}\ ; \nonumber
 \end{eqnarray}

\noindent and

 \begin{equation}
 \label{MCS}
 S^{CS}= T_0\int dt\, {\rm STr} \{ -P[k^{-2} k^{(1)}]+iP[(i_X i_X)C^{(3)}] +
 \frac{1}{2} P[(i_X i_X)^2 i_k C^{(6)}] +\dots\}\ ,
 \end{equation}

\noindent where the dots include couplings to higher order background potentials and
 products of  different background fields contracted with the non-Abelian
 scalars\footnote{These couplings are not shown explicitly because they will
 not play a role in the backgrounds under consideration in this paper.}. 

In (\ref{MBI})+(\ref{MCS}) $k^\mu$ is the
Abelian Killing vector which, by construction (see \cite{JL2}), points on the direction of
propagation of the waves. This direction is isometric\footnote{Indeed, the
  action (\ref{MBI})+(\ref{MCS}) is a gauged $\sigma$-model in the spirit of
  \cite{O}, \cite{BJO}, \cite{BLO}, \cite{EJL} , which eliminates the dependence in the isometric coordinate.},
because the background fields are either contracted with the Killing vector,
so that any component along the isometric direction of the contracted field
vanishes, or pulled back in the worldvolume with covariant derivatives
relative to the isometry (see \cite{JL2} for their explicit
definition)\footnote{The reduced metric ${\cal G}_{\mu\nu}$ appearing in
  (\ref{mred}) is in fact defined such that its pull-back with ordinary
  derivatives equals the pull-back of $g_{\mu\nu}$ with these covariant
  derivatives.}. Therefore, the action exhibits a $U(1)$ isometry associated to
translations along this direction, which, by construction, is also the direction
on which the waves propagate. For more details about the construction of the
action, see \cite{JL1} and \cite{JL2}.

 As we have pointed, the action (\ref{MBI}) $+$ (\ref{MCS}) has been successfully used in the
 microscopical study of giant graviton configurations in backgrounds which are
 not linear perturbations to Minkowski.
In all cases perfect agreement with the
 description of \cite{giants} has been found in the limit of large number of
 gravitons, in  which the commutative configurations of \cite{giants} become
 an increasingly better approximation to the non-commutative microscopical
 configurations, in very much the same spirit as in \cite{M}. Although a number of subtleties can be risen (involving \textit{e.g.} the symmetrized trace or even the proper form of the action as a low energy effective action), we will take a more pragmatical point of view, and use the action in very much the same spirit as in \cite{JLR1}, \cite{JLR2}, \cite{JLR3}, \cite{LR}.

\section{The extremal black string in M-theory on $T^6$}

We will concentrate in a certain class of compactifications of M-theory on
$T^6$. We will parametrize the 6-torus with $\{y^i,\ i=1,\dots 6\}$. By
placing three stacks of M5 branes with charges $\{p_1,p_2,p_3\}$ carrying
momentum $q_0$ along a direction $y$ and wrapping respectively
$\{y,y^3,y^4,y^5,y^6\}$, $\{y,y^1,y^2,y^5,y^6\}$ and $\{y,y^1,y^2,y^3,y^4\}$,
we have that the metric is given by

\begin{equation}
\label{Mmetric}
ds^2=h^{-\frac{1}{3}}\Big[-dt^2+dy^2+\frac{q_0}{r}(dt-dy)^2\Big]+h^{\frac{2}{3}}\Big[dr^2+r^2\big(d\theta^2+\sin^2\theta
  d\phi^2\big)\Big]+h^{-\frac{1}{3}}\Sigma_{i=1,2,3}H_ids^2_{T_i}\ .
\end{equation}

There is also a 3-form potential, given by

\begin{equation}
\label{Mform}
C^{(3)}=\sin\theta d\theta\wedge
d\phi\wedge\Big[p_3\frac{y^5dy^6-y^6dy^5}{2}+p_2\frac{y^3dy^4-y^4dy^3}{2}+p_1\frac{y^1dy^2-y^2dy^1}{2}\Big]\ .
\end{equation}

We have defined

\begin{center}
\begin{tabular}{c c c}
$h=H_1H_2H_3\ ;$&$H_i=1+\frac{p_i}{r},\ i=1,2,3\ ;$&$H_0=1+\frac{q_0}{r}\ .$
\end{tabular}
\end{center}

The charges $p_i$ of the stacks of branes are related to the number of branes
$n_i$ as

\begin{equation}
p_i=\frac{2\pi^2n_i}{M_{11}^3 vol(T_i)}\ .
\end{equation}

The geometry given by (\ref{Mmetric}) and (\ref{Mform}) is that of a
black string, whose horizon is the $S^2$.

We can take the near horizon limit of this geometry. The way in which this limit is taken is somehow
subtle, since it is different depending whether $q_0$ zero is or not (the
details can be seen in, for example, \cite{DGMSWZ}). However, the
resulting space is in both cases $AdS_3\times S^2\times T^6$, whose metric in
global coordinates is

\begin{equation}
\label{ads3metricM}
ds^2=R^2[-\cosh^2\chi d\tau^2+d\chi^2+\sinh^2\chi
  d\varphi^2]+\tilde{R}^2d\Omega_2+ds^2_{T^6}(y^i)\ .
\end{equation}

There is also a 3-form potential, given by

\begin{equation}
\label{ads3formM}
C^{(3)}=A_i(y^j)d\omega_2\wedge dy^i\ .
\end{equation}

\noindent Here $d\omega_2$ in (\ref{ads3formM}) stands for the volume form in
the 2-sphere.

The parameters $R$ and $\tilde{R}$ are given in terms of
$\lambda=(p_1p_2p_3)^{\frac{1}{3}}$ as

\begin{center}
\begin{tabular}{c c}
$R=2\lambda\ ,$ & $\tilde{R}^2=\frac{1}{\lambda}\ .$
\end{tabular}
\end{center}

\subsection{M2 wrapping the horizon}

We will consider an M2 brane wrapping the $S^2$ and moving in the $\varphi$
direction in the background given by
(\ref{ads3metricM})+(\ref{ads3formM}). We will also take the ansatz
that it lives at $y^i=0$. Then, by using the standard $DBI+CS$ effective
action for the brane, it is straightforward to see that the
action for such a configuration is

\begin{equation}
\label{SM2}
S=-4\pi T_2\tilde{R}^2R\int d\tau
\sqrt{\cosh^2\chi-\sinh^2\chi\dot{\varphi}^2}\ ,
\end{equation}

\noindent where $T_2$ stands for the M2 tension.

Since $\varphi$ does not appear in the action, its canonically conjugated
momentum is a conserved quantity. Performing a Legendre transformation, we can
switch to the conserved hamiltonian

\begin{equation}
\label{HM2}
H=\cosh\chi\sqrt{16\pi^2T_2^2\tilde{R}^4R^2+\frac{P^2}{\sinh^2\chi}}\ .
\end{equation} 

Minimizing (\ref{HM2}) with respect to $\chi$ we find a minimum at 

\begin{equation}
\label{HM2minimumposition}
\sinh^2\chi=\frac{P}{4\pi T_2\tilde{R}^2R}\ .
\end{equation}

\noindent For this minimum, the energy of the configuration is

\begin{equation}
\label{HM2minimum}
E=P+4\pi T_2\tilde{R}^2R\ .
\end{equation}

\noindent This particular dispersion relation can be related with
the particular isometry group of the $AdS_3$ geometry, as can be seen in
\cite{DGMSWZ}.

This configuration is a particular case of the ones considered in
\cite{DGMSWZ}, when we assume the brane to be at the origin
of the $T^6$. Indeed, the results obtained here coincide with the ones in
that reference once we assume that there is no velocity in $\mathcal{M}$ (we
will go back to this point later). It is important to notice that due to this ansatz there is no contribution from the $CS$ part of
the action, since in the ansatz $y^i=0$ the would-be coupling to the 3-form vanishes. Then,
the geometry is stable only due to the metric background. Notice that if the
brane had zero momentum in $\varphi$, it would be sitting in the center of the
$AdS$, as one can see
from (\ref{HM2minimumposition}). In a sense, the ``centrifugal force''
is pulling the configuration out of the center against the $AdS$ throat.

\subsection{Microscopical description of the M2 wrapping the horizon}

In order to give a microscopical description of the M2 brane wrapping the
horizon of the previous subsection, we will use the action for 11 dimensional
coincident gravitons. 

Since we are taking the configuration to be at the
origin of the torus, we see that there is no contribution from the 3-form
potential, which vanishes for constant $y^i$. Therefore the only
contribution to the action will be that of the $DBI$, given by (\ref{MBI}), in
which the coupling to the 3-form potential is zero.

Writing the effective background in suitable cartesian coordinates it reads

\begin{equation}
\label{ads3Meff}
ds^2=R^2[-\cosh^2\chi d\tau^2+d\chi^2+\sinh^2
  d\varphi^2]+\tilde{R}^2(dx^i)^2\ ,
\end{equation}

\noindent where $\{x^i\ |\ \Sigma (x^i)^2=1\}$ are the cartesian coordinates of the $S^2$.

Since our macroscopic M2 was moving in the $\varphi$ direction, we should
take in the microscopic description $\varphi$ as the direction of propagation
of the gravitational waves, \textit{i.e.}
$k^{\mu}=\delta^{\mu}_{\varphi}\rightarrow k^2=R^2\sinh^2\chi$.

The fuzzy manifold to which the gravitons will expand will be the $S^2$, for
which we will take the non-commutative $S^2$ ansatz

\begin{equation}
\label{ncansatzM}
X^i=\frac{1}{\sqrt{C_2}}J^i\ ,
\end{equation}

\noindent where $J^i$ are the $SU(2)$ generators in an N-dimensional
fundamental representation of $SU(2)$ whose quadratic Casimir is
$C_2$. Therefore they satisfy $\Sigma (X^i)^2=1$ as a matrix identity.

By substituting this ansatz into (\ref{MBI}) we have that the $Q$ matrix in
(\ref{MBI}) is

\begin{equation}
Q^i_j=\delta^i_j+i\tilde{R}^2R\sinh\chi [X^i,X^j]\ .
\end{equation}

It is easy to see that the piece in (\ref{MBI}) involving $(Q^{-1}-1)$ does
not contribute, and therefore the only purely non-abelian contribution is that
of $det(Q)$. Whith the help of the symmetrized trace it can be seen that

\begin{equation}
det(Q)=1+\frac{4\tilde{R}^4R^2\sinh^2\chi}{C_2}\ .
\end{equation}

Since the configuration is effectively static, we can compute the hamiltonian
easily as $H=-L$. In addition, expanding the square root in the Myers limit of
\cite{M} $\tilde{R}\ll \sqrt{N}$, we can compute the symmetrized trace and regard
the expression as the expansion of

\begin{equation}
\label{hm2}
H=\cosh\chi\sqrt{\frac{4N^2T_0^2\tilde{R}^4R^2}{C_2}+\frac{(NT_0)^2}{\sinh^2\chi}}\
\end{equation}

\noindent up to higher order corrections in $\frac{R}{N^2}$.

Taking into account that in our conventions $T_0=2\pi T_2$, and that by
construction $P=NT_0$, it is clear that in the large $N$ limit when
$\frac{N}{C_2}\rightarrow 1$ both the expressions (\ref{HM2}) and (\ref{hm2})
coincide. 

Thus, we see that there is a microscopical description for the M2 brane
wrapping the horizon in terms of dielectric gravitons expanded due to
dielectric effect. This expansion is caused not by the flux, but by the proper
geometry, and therefore it is a nice example of purely gravitational dielectric
effect. 

Extremizing (\ref{hm2}) we find that there is a minimum at

\begin{equation}
\label{hm2minimumposition}
\sinh^2\chi=\frac{\sqrt{C_2}}{2\tilde{R}^2R}\ ,
\end{equation}

\noindent whose energy is

\begin{equation}
\label{hm2minimum}
E=NT_0+2\tilde{R}^2RT_0\frac{N}{\sqrt{C_2}}\ .
\end{equation}

In large $N$ both the position and the value of the energy at the minimum
coincide with their macroscopical counterparts given by
(\ref{HM2minimumposition}) and (\ref{hm2minimum}). This is to be expected,
since the description in terms of gravitational waves is only valid for
coincident GW. By taking $N$ large enough this regimen of validity overlaps
with that of the macroscopical description, exactly as it was discussed in \cite{M}.

It is worth saying a few words on the stability of our solution. On one hand it is wrapping a non-contractible cycle in the background geometry, and therefore it is topologically stable against contraction to a point. Contrary to the giant graviton case, where the radius of the configuration is a modulus which we should put on-shell, now the radius of the configuration is simply the $AdS$ radius, which is fixed. On the other hand, as we will show later, the configuration is in a Lowest Landau Level in the internal torus, which means that it is already in the ground state and therefore cannot decay. The only possible instability would be in the position in the $AdS$, but simply by plotting the energy we see that there is no other extrema for our system that the one found (which is an absolute minimum). Therefore it is clear that the configuration is stable. As we will see, in the IIA case the position of the configuration in the $AdS$ can be even mapped to the condition for the configuration to be one half BPS, which is again pointing to the stability of the configuration. 

\section{From black strings in M-theory to black holes in IIA}

The geometry given by (\ref{Mmetric})+(\ref{Mform}) can be reduced to a type
IIA black hole by performing a Kaluza-Klein reduction along the $y$
direction. The resulting metric is

\begin{equation}
\label{IIAmetric}
ds^2=-(H_0h)^{-\frac{1}{2}}dt^2+(H_0h)^{\frac{1}{2}}\Big[dr^2+r^2\big(d\theta^2+\sin^2\theta
  d\phi^2\big)\Big]+(\frac{h}{H_0})^{-\frac{1}{2}}\Sigma_{i=1,2,3}H_ids^2_{T_i}\ .
\end{equation}

There is also a 3-form RR potential given by

\begin{equation}
\label{IIA3form}
C^{(3)}=\sin\theta d\theta\wedge
d\phi\wedge\Big[p_3\frac{y^5dy^6-y^6dy^5}{2}+p_2\frac{y^3dy^4-y^4dy^3}{2}+p_1\frac{y^1dy^2-y^2dy^1}{2}\Big]\ .
\end{equation}

Since in 11 dimensions the 3 stacks of M5 branes carry momentum along a
worldvolume direction $y$ (which is the one along we reduce), once we reduce
we have 4 intersecting stacks of D4 branes which carry a non-zero D0-brane charge (indeed, one can see that if
$q_0=0$ there is a null singularity at $r=0$ in IIA) given by a RR 1-form
potential 

\begin{equation}
\label{IIA1form}
C^{(1)}=(1-\frac{1}{H_0})dt\ .
\end{equation}

There is also a non-zero dilaton, given by

\begin{equation}
\label{IIAdilaton}
e^{\Phi}=\frac{H_0^3}{h}\ .
\end{equation}

Again, there is a horizon in $r=0$. We can take the near horizon limit of the
IIA geometry to have\footnote{We also change to global coordinates. The
  details can be found in \cite{DGMSWZ}.}

\begin{eqnarray}
\label{IIAadsmetric}
ds^2&=&R_{IIA}^2\Big[-\cosh^2\chi d\tau^2+d\chi^2\Big]+R_{IIA}^2\Big[d\theta^2+\sin^2\theta
  d\phi^2\Big]+\\ \nonumber
  &&\sqrt{\frac{q_0p_1}{p_2p_3}}\Big((dy^1)^2+(dy^2)^2\Big)+\sqrt{\frac{q_0p_2}{p_1p_3}}\Big((dy^3)^2+(dy^4)^2\Big)+\sqrt{\frac{q_0p_3}{p_2p_1}}\Big((dy^5)^2+(dy^6)^2\Big)\ ,
\end{eqnarray}

\begin{equation}
\label{IIAads3form}
C^{(3)}=\sin\theta d\theta\wedge
d\phi\wedge\Big[p_3\frac{y^5dy^6-y^6dy^5}{2}+p_2\frac{y^3dy^4-y^4dy^3}{2}+p_1\frac{y^1dy^2-y^2dy^1}{2}\Big]\ ,
\end{equation}

\begin{equation}
\label{IIAads1form}
C^{(1)}=-\frac{R_{IIA}^2}{q_0}(1-\sinh\chi)d\tau\ ,
\end{equation}

\begin{equation}
\label{IIAadsdilaton}
e^{\Phi}=\frac{q_0}{R_{IIA}}\ .
\end{equation}

The quantity $R_{IIA}$ appearing in the IIA near horizon limit is defined as
$R_{IIA}=(q_0p_1p_2p_3)^{\frac{1}{4}}$.

The geometry given by
(\ref{IIAadsmetric})+(\ref{IIAads1form})+(\ref{IIAads3form})+(\ref{IIAadsdilaton})
is that of $AdS_2\times S^2\times T^6$ in global coordinates. Note that it
cannot be obtained as a reduction of the 11 dimensional near horizon
geometry. 

\subsection{D0 branes in the IIA $AdS_2\times S^2\times T^6$ geometry}

We have obtained the IIA $AdS_2$ as a dimensional reduction of the
corresponding background of M-theory. As we have so far seen, in 11 dimensions
there are stable configurations of gravitons expanded to an $S^2$ in the near
horizon limit which are moving along a cycle contained in the $AdS_3$. When
regarded from the point of view of the full geometry instead of just the
near-horizon limit, the configuration of gravitons inherits momentum in the
$y$ direction. Since we are reducing precisely in this coordinate, we should
expect to have stable static configurations of D0 branes wrapping the $S^2$ as counterparts of
the M-theory construction\footnote{Remember we are assuming static
  configurations in $\mathcal{M}$ through all the paper. Therefore there will
  be no coupling to the background potential.}.

\subsubsection{M-theory gravitons in the full geometry}

Before analyzing the IIA configurations of coincident D0 branes, we will go
back to 11 dimensions and analyze coincident gravitons in the full geometry
given by (\ref{Mmetric})+(\ref{Mform}). 

In order to understand the IIA configurations from 11d, assume now the
particular configuration in which the gravitational waves carry momentum only
in the $y$ direction\footnote{In general there could be momentum in other
  coordinates, but since we will reduce along $y$, it will suffice for our purposes with
  considering momentum in $y$.},
\textit{i.e} $k^{\mu}=\delta^{\mu}_y$. Taking the same ansatz as before,
it is straightforward to see that the energy for this type of configuration
is

\begin{equation}
\label{HMfull}
E=\frac{NT_0}{1+\frac{q_0}{r}}\Big\{\sqrt{1-\frac{q_0^2}{r^2}}\sqrt{1+\frac{4r^2h}{C_2}(1+\frac{q_0}{r})}+\frac{q_0}{r}\Big\}\ .
\end{equation}

\noindent The last term in (\ref{HMfull}) comes from the monopole term in
(\ref{MCS}). The monopole coupling in the action for gravitational waves
represents the fact that gravitons are fundamentally charged with respect to a
momentum operator. Since this momentum operator involves a covariant
derivative\footnote{Remember that it reads
  $k^{-2}k_{\mu}\mathcal{D}X^{\mu}$.}in the gauged $\sigma$-model which is the
action for gravitational waves, it only contributes when $k_0+k_i\partial X^i$
(being $i$ different from the direction of propagation) is non-zero. In our
particular set-up, $\partial X^i=0$, but $k_0=g_{0y}\ne 0$, and therefore
there is a non-zero contribution from the monopole term. Notice that once we
reduce along $y$, this monopole coupling to a momentum operator in 11
dimensions gets mapped to a monopole coupling to $C^{(1)}$ in IIA, as expected,
since when reducing from M-theory to IIA momentum in the 11th direction
becomes D0-brane charge. From this discussion we see that there should exist
similar configurations of expanded D0 branes in IIA, which should be now
static.

The expression (\ref{HMfull}) is rather complicated as a function of $r$. In any case, since the
energy must be real, we have that $r\ge q_0$. In general, for generic
values of $p_1,\ p_2,\ p_3,\ q_0$, we will have stable expanded branes away
from the origin.

Interestingly, also in the full geometry we have a configuration of
expanded gravitational waves which is stable only due to momentum, with no
need of the form potentials. This means that also in the full geometry we have
a purely gravitational dielectric effect. Since in the full geometry the waves
are moving in the $y$ direction, which is precisely the one along which we are
reducing, we should expect analogue configurations in type IIA, this time 
in
terms of D0 branes. These IIA configurations should be stable not due to 
3-form
RR potential, but due to 1-form potential, \textit{i.e.} the mechanism which
renders the IIA configuration stable should be the gravitational dielectric
effect but seen with KK glasses.

\subsubsection{D0 branes in type IIA}

Working in the near-horizon limit of the full type IIA geometry, we will
consider $N$ coincident D0 branes living at the origin of the
6-torus. Therefore we will take $y^i=0$. Then, there is no coupling of the
3-form RR potential. The effective background for the D0 branes in suitable
cartesian coordinates for the $S^2$ is

\begin{equation}
ds^2=R_{IIA}^2\Big(-\cosh^2\chi
d\tau^2+d\chi^2\Big)+R_{IIA}^2\Big(dx^2+dy^2+dz^2\Big)\ ,
\end{equation}

\noindent and

\begin{equation}
C^{(1)}_{\tau}=-\frac{R_{IIA}^2}{q_0}(1-\sinh\chi)\ .
\end{equation}

Since $\Sigma (x^i)^2=1$, we will take the same non-commutative ansatz for the
$x^i\rightarrow X^i$ as (\ref{ncansatzM}). Then, upon particularizing the
action for coincident D0 branes given in \cite{M}, and taking the suitable limit as we did in the M-theory case, we have that the action for
the system has a DBI and a CS part which read

\begin{equation}
S_{BDI}=-\frac{NT_0R_{IIA}^2}{q_0}\int
d\tau\cosh\chi\sqrt{1+\frac{4R_{IIA}^4}{C_2}}\ ,
\end{equation}

\begin{equation}
S_{CS}=-\frac{NT_0R_{IIA}^2}{q_0}\int d\tau(1-\sinh\chi)\ .
\end{equation}

\noindent Notice that the last term in (\ref{HMfull}) agrees exactly with
$S_{CS}$ once we take the near horizon limit. This reflects the anticipated
fact that the IIA construction is the KK version of the gravitational
dielectric effect.

Since the configuration is static, $H=-L=-L_{DBI}-L_{CS}$:

\begin{equation}
\label{HD0}
E=\frac{NT_0R_{IIA}^2}{q_0}\Big\{\cosh\chi\sqrt{1+\frac{4R_{IIA}^4}{C_2}}-\sinh\chi+1\Big\}\ .
\end{equation}

\noindent Minimizing (\ref{HD0}) with respect to $\chi$ we have a minimum at

\begin{equation}
\label{D0minimumposition}
\tanh\chi=\frac{1}{\sqrt{1+\frac{4R_{IIA}^4}{C_2}}}\ ,
\end{equation}

\noindent whose energy is 

\begin{equation}
\label{Eminimum}
E=\frac{NT_0R_{IIA}^2}{q_0}\Big\{\frac{2R_{IIA}^2}{\sqrt{C_2}}+1\Big\}=\frac{NT_0R_{IIA}^2}{q_0}+\frac{2T_0R_{IIA}^4}{q_0}\frac{N}{\sqrt{C_2}}\ .
\end{equation}

\subsubsection{D2 brane as effective description}

We can have an alternative description of the configuration in the previous
section in terms of a D2 brane wrapping the horizon. This has been already
presented in \cite{DGMSWZ}. We will take static
gauge, and assume the brane to be at the origin of the torus. Since the D2 is
spherical, it carries no net D2-brane charge. In order to give the 
suitable D0
brane charge we have to turn on a non-zero DBI vector field, whose field
strength is

\begin{equation}
F_{\theta \phi}=\frac{N}{2}\sin\theta\ ,
\end{equation}

\noindent where $N$ is the total D0 brane charge which the D2 carries.

By using the standard DBI+CS action for a D2 brane in the background we have,
it is straightforward to see that the energy for the system (which is $-L$
since the configuration is static) is:

\begin{equation}
\label{HD2}
E=\frac{2\pi
  T_2NR_{IIA}^2}{q_0}\Big\{\cosh\chi\sqrt{1+\frac{4R_{IIA}^4}{N^2}}-\sinh\chi+1\Big\}\ .
\end{equation}

\noindent Minimizing with respect to $\chi$ we have a minimum located at

\begin{equation}
\label{HD2minimumposition}
\tanh\chi=\frac{1}{\sqrt{1+\frac{4R_{IIA}^4}{N^2}}}\ .
\end{equation}

\noindent Indeed this condition is precisely the one found in \cite{SSTY} for a supersymmetric D2 brane with DBI flux on it. This means that our microscopical construction is one half BPS at least in the large $N$ limit, when it overlaps with its macroscopical counterpart. In some sense this implyies the stability of the construction, since on one hand it is clear that it its topologically stable (it wraps a non-contractible cycle in the space), and on the other hand, as we will show later, it is in its ground state in the internal torus (it is a Lowest Landau Level). Therefore it cannot decay. The only possible instability would be that the location in the $AdS$ was not stable, but SUSY ensures now the stability. In any case it is easy to see that there is no other minimum in $\chi$ to which the brane could jump (for example simply by plotting the energy one can see immediately that the only extremum is the absolute minimum we found). Therefore we conclude that the configuration is stable. 

The energy for this minimum is:

\begin{equation}
\label{HD2minimum}
E=\frac{2\pi T_2NR_{IIA}^2}{q_0}+\frac{4\pi T_2R_{IIA}^4}{q_0}\ .
\end{equation}

\noindent Once we take into account that $2\pi T_2=T_0$ it is straightforward to see
that (\ref{HD2}), (\ref{HD2minimumposition}) and (\ref{HD2minimum}) coincide
in the large $N$ limit (when in the microscopical computation
$\frac{N}{\sqrt{C_2}}=1$) with their microscopical counterparts given by
(\ref{HD0}), (\ref{D0minimumposition}) and (\ref{Eminimum}). This coincidence
is to be expected precisely only in the large $N$ limit, since it is only
there when we have that both descriptions are valid at the same time, exactly
as we have pointed out when discussing the M-theory configuration.

\section{IIB black holes}

So far we have seen that there is a stable M2 brane wrapping the horizon of
the 11 dimensional black string, which can be described microscopically in
terms of gravitational waves expanded due to dielectric effect. This
dielectric effect is due to purely gravitational effects, since there is no
contribution from the 3-form potential. Upon reduction along a direction of
propagation, which is related to the one in which the M-theoretic configuration
is moving, this gives rise to a type IIA configuration of expanded D0 branes
in which there is no dielectric coupling to a 3-form potential. Its stability
is only due to gravitational effects plus the monopolar coupling of each D0
brane to the background 1-form potential. This is a consequence of the fact
that the original configuration in 11 dimensions is static only due to
gravitational effects. 

In this section we will see another example of configuration of brane wrapping
the horizon (which has also been studied from the macroscopical point of view in \cite{DGMSWZ}), this time in type IIB, whose dispersion relation is of the form
of the ones we have so far worked out, and which admits a
microscopical description in terms of gravitational waves expanded again due
to purely gravitational dielectric effect. 

We will focus on the type IIB $AdS_3\times S^3\times T^4$ background obtained
by taking type IIB string theory in $T^4\times S^1$ and wrapping D1 strings on
the $S^1$ and $D5$ branes on the $T^4\times S^1$. Once we take the near horizon limit we end up with the following
$AdS_3\times S^3\times T^4$ background (indeed, we will work with a 2 times
T-dualized along the torus coordinates $y^1,\ y^2$ version of the background, which is then the near horizon geometry
of two sets of D3 branes. For further details see \cite{DGMSWZ}):

\begin{equation}
\label{IIBmetric}
ds^2=L^2\Big(-\cosh^2\chi d\tau^2+d\chi^2+\sinh^2\chi
d\varphi^2\Big)+L^2d\Omega_3^2+ds_{T^4}^2\ .
\end{equation}

In this background there is no dilaton. In addition, the radius of the $AdS$,
$L$, can be written in terms of the charges of the background. Namely $L^2=\sqrt{Q_1Q_2}$.

There is also a 4-form potential which has entries on the $T^4$, which
schematically reads

\begin{equation}
\label{C4}
C^{(4)}=\Big\{F_1d\tau\wedge d\varphi+F_2d\psi\wedge d\phi+F_3d\tau\wedge
d\psi+F_4d\varphi\wedge d\phi\Big\}\wedge dy^1\wedge dy^2\ .
\end{equation}

Since we will be again taking the ansatz that the configuration is static in
$T^4$, we see that there will be no coupling to the 4-form RR
potential\footnote{In any case, from (\ref{C4}), one can see that the 
4-form
  does not couple to the D3 brane worldvolume theory, since the RR would be
  coupling would be
$P[C^{(4)}]=\partial_{i}x^{\nu_1}\partial_{j}x^{\nu_2}\partial_{k}x^{\nu_3}\partial_{l}x^{\nu_4}C_{\nu_1\nu_2\nu_3\nu_4}\epsilon^{ijkl}$.
Then, it is easy to see that this gives no contribution with the
particular form potential in (\ref{C4}).}. We are thus left only with the
metric background given by (\ref{IIBmetric}).

\subsection{The action for IIB gravitational waves}

In order to have an action for IIB coincident waves we have to perform a dimensional reduction plus
a T-duality from the M-theory one given by (\ref{MBI})+(\ref{MCS}) down to IIB. This introduces a new isometry in
the action, which will have an associated abelian Killing vector
$l^{\mu}$ pointing in the direction along which we performed the T-duality
(say $z$). Since in the background we are considering we have a 4-form
potential, we will keep non-vanishing couplings to $C^{(4)}$. Indeed, the would-be coupling to the 4-form is achieved by contracting
$C^{(4)}$ with $l^{\mu}$. We see therefore that the extra isometry is playing
a key role. The final type IIB action with coupling to the
4-form potential is given by\footnote{For further details about the construction of this
action, see \cite{JLR1}.}:

\begin{equation}
\label{IIBBI}
S^{BI}=-T_0\int d\tau\ STr\Big\{ k^{-1}
\sqrt{-P[E_{00}+E_{0i}(Q^{-1}-\delta)^i_k E^{kj}E_{j0}]
det(Q^i_j)}\Big\}\ ,
\end{equation}

\noindent where now

\begin{eqnarray}
E_{\mu\nu} &=& g_{\mu\nu}-k^{-2}k_\mu k_\nu -l^{-2}l_\mu l_\nu
-k^{-1}l^{-1} e^{\phi}(i_ki_l C^{(4)})_{\mu\nu} \\
Q^i_j &=& \delta^i_j + i[X^i,X^k]e^{-\phi} k l E_{kj}
\end{eqnarray}

\noindent Here $k^{\mu}$ is the Killing vector pointing along the direction of propagation of
the gravitons, $k^2=g_{\mu\nu}k^\mu k^\nu$, $k_\mu=g_{\mu\nu}k^\mu$, and
the same notation applies to $l^\mu$. Here we have also taken
$g_{\mu\nu}k^\mu l^\nu=0$, a condition that is satisfied for the background
that we consider in this paper.

The Chern-Simons part of the action contains the term \cite{JLR1}

\begin{equation}
\label{IIBCS}
S_{CS}=T_0 \int P[(i_Xi_X)i_lC^{(4)}]
\end{equation}

\noindent in addition to the usual monopole coupling to a momentum operator, exactly as
the first term in (\ref{MCS}).

The fact that the direction of propagation
appears as an isometric direction is common to all gravitons in Type II and
M theories (see \cite{JL2}, \cite{JLR1}). The second isometric direction, $z$, is however
special to the Type IIB case (see \cite{JLR2} for further details about the action). It is remarkable that only due to the presence of this isometric
direction we can obtain dielectric couplings to higher order Type IIB RR
potentials.

\subsection{Dielectric gravitons wrapping the horizon in $AdS_3\times
  S^3\times T^4$}

Using the action (\ref{IIBBI})+(\ref{IIBCS}), we will consider dielectric
gravitons wrapping the horizon in the $AdS_3\times S^3\times T^4$
geometry. As before, we will take the ansatz that we are at the origin of the
torus. We will also assume that our configurations are moving in the $\varphi$
direction of the background (\ref{IIBmetric}). Therefore we will take in the
non-abelian action for gravitons $k^{\mu}=\delta^{\mu}_{\varphi}$.

In the action (\ref{IIBBI})+(\ref{IIBCS}) there is an extra isometry with
respect to the 11d action (\ref{MBI})+(\ref{MCS}), which is due to the
neccesary T-duality direction to arrive to a IIB construction. At first sight,
once we assume the configuration to be in the origin of the $T^4$, the
background (\ref{IIBmetric}) does not have any other $U(1)$ isometry but
$\varphi$. Nevertheless, the $S^3$ can be regarded as the Hopf fibering
$S^1\rightarrow S^2$. Then we see that there is an internal isometry hidden in
a non-trivial way inside the 3-sphere. Going to adapted coordinates to see
this fibering structure we have that the background reads\footnote{Again, we
  will omit the pieces living in the $T^4$.}:

\begin{equation}
\label{IIBmetricisometry}
ds^2=L^2\Big(-\cosh^2\chi d\tau^2+d\chi^2+\sinh^2\chi
d\varphi^2\Big)+(\frac{L}{2})^2\Big(dx^2+dy^2+dz^2\Big)+(\frac{L}{2})^2(d\chi_3+A)^2\ ,
\end{equation}

\noindent where we have written the $S^2$ base in cartesian coordinates for
latter convenience. 

The $A$ connection in (\ref{IIBmetricisometry}) gives the necesary twist to the
$U(1)$ fiber parametrized with $\chi_3$ to have an $S^3$. It can be explicitly
seen in \cite{JLR1}.

It is now natural to identify $l^{\mu}=\delta^{\mu}_{\chi_3}$. Therefore the
non-commutative manifold will be not the full $S^3$ but the $S^2$ base. This
construction is very similar to that in \cite{JLR1}. While in that reference
it was used to describe microscopically the $S^3$ giant graviton which lives
in $AdS_5\times S^5$, here we use it to describe microscopically a
configuration wrapping the $S^3$ horizon in $AdS_3\times S^3\times T^4$. In this case, although we are
working with the same metric background as in \cite{JLR2}, we need to
construct a fuzzy $S^3$, which is very much the same
situation as in \cite{JLR1}. 

Taking as non-commutative anstaz the same as (\ref{ncansatzM})

\begin{equation}
\label{ncansatzIIB}
X^i=\frac{1}{\sqrt{C_2}}J^i
\end{equation}

\noindent it is straightforward to see that the hamiltonian for this
configuration (again computed as minus the lagrangian) reads:

\begin{equation}
\label{HIIBm}
H=NT_0\frac{\cosh\chi}{\sinh\chi}\sqrt{1+\frac{(Q_1Q_2)^2}{16C_2}\sinh^2\chi}\ .\end{equation}

\noindent Here we have made use of the symmetrized trace and taken the
suitable large $N$ limit as in the previous cases.

Minimizing with respect to $\chi$ we see a minimum at

\begin{equation}
\label{HIIBmminimumposition}
\sinh^2\chi=\frac{4\sqrt{C_2}}{Q_1Q_2}\ ,
\end{equation}

\noindent whose energy is

\begin{equation}
\label{HIIBmminimum}
E=NT_0+\frac{N}{4\sqrt{C_2}}T_0Q_1Q_2.
\end{equation}

\subsection{D3 brane wrapping the horizon in $AdS_3\times S^3\times T^4$}

Exactly as we did before, we can have an effective description of this
configuration in terms of a D3 brane wrapping the horizon if we take
sufficiently large $N$ in the microscopical configuration.

Assuming static gauge for a D3 brane sitting at the origin of the torus and
wrapping the $S^3$ with non-zero velocity in the $\varphi$ direction of
(\ref{IIBmetric}) it is easy to see that the action is:

\begin{equation}
\label{D3S}
S=-T_3L^4\Omega_3\int d\tau\ \sqrt{\cosh^2\chi-\sinh^2\chi\dot{\varphi}^2}\ .
\end{equation}

\noindent Here $\Omega_3$ stands for the volume of a unit 3-sphere, while $T_3$ is the
tension of the 3-brane, whose relation with $T_0$ can be seen in \cite{JLR1}.

Since $\varphi$ is a cyclic coordinate, we can switch to hamiltonian formalism
in terms of its canonically conjugated and conserved momentum $P$. The energy
reads:

\begin{equation}
\label{HD3M}
E=P\frac{\cosh\chi}{\sinh\chi}\sqrt{1+\frac{T_0^2(Q_1Q_2)^2}{16P^2}\sinh^2\chi}\ .
\end{equation}

Comparing (\ref{HD3M}) with (\ref{HIIBm}) we see that there is a perfect
agreement once we take into account that $NT_0=P$ by construction and that in
the large $N$ limit when both descriptions are expected to coincide $N\sim
\sqrt{C_2}$.

Minimizing (\ref{HD3M}), we get a minimum located at

\begin{equation}
\label{HD3Mminimumposition}
\sinh^2\chi=\frac{4P}{T_0Q_1Q_2}\ ,
\end{equation}

\noindent for which the corresponding energy is

\begin{equation}
\label{HD3Mminimum}
E=P+\frac{T_0}{4}Q_1Q_2\ .
\end{equation}

As we see, in the large $N$ limit, the macroscopical expressions coincide with
their microscopical counterparts.

Again, since our configurations wrap non-contractible cycles, they are stable against collapse to a point. Since one can show that the energy has only one extremum (which is an absolute minimum) where our configurations sit, we conclude that they must be stable.

\section{On the ansatz of zero momentum in the internal manifold}

So far we have been considering the ansatz that our configurations are static
in the internal manifold (\textit{i.e.} the torus). We will return now to
this point to see explicitly how this is a consistent ansatz. For this purpose we note that the
general form of the action for both the M-theory and IIA macroscopic\footnote{For the sake of
  simplicity, we will think in terms of the macroscopic description, although
  we could equally consider it from the microscopic point of view.}
configurations if we allow the brane to move in the torus (whose coordinates
are generically denoted with $y$) is 

\begin{equation}
\label{L}
S=\int d\tau\ L=\int d\tau\ \big\{-F\sqrt{G-\dot{y}^2}+C(y)\dot{y}\big\}\ ,
\end{equation}

\noindent where $F$ and $G$ are the suitable functions in each particular case
of the charges and (if in type IIB or M-theory) the velocity in the cycle
contained in the $AdS$; and $C(y)$ stands for the appropriate function coming
from the coupling to the form potential, which in general could depend on the
torus coordinates. Notice that we have integrated the dependence in the volume
of the brane, and thus the action is analogous to the action for a charged
particle in a magnetic field.

If we vary this action with respect to $y$ we have that the equations of
motion are of the form

\begin{equation}
\label{eom}
\frac{\partial C(y)}{\partial y}\dot{y}+\frac{\partial}{\partial\tau}\Big(
\frac{F\dot{y}}{\sqrt{G-\dot{y}^2}}-C(y)\Big)=0\ .
\end{equation}

From (\ref{eom}) we see that once we assume $y=constant$ the equations of
motion are trivially satisfied (recall that $C(y=const)=constant$), showing that our ansatz is
consistent. Furthermore, we can compute the conjugated momentum to $\dot{y}$,
which reads

\begin{equation}
\label{mom}
\Pi=\frac{F\dot{y}}{\sqrt{G-\dot{y}^2}}-C(y)\ .
\end{equation}

\noindent If we replace the ansatz of constant $y$ we have that 

\begin{equation}
\label{momenansatz}
\Pi=-C
\end{equation}

But as can be seen in \cite{DGMSWZ}, this identification is like restricting
ourselves to the lowest energy level, which is indeed in a Lowest Landau Level, as was pointed out in (\cite{DGMSWZ}).

\section{Conclusions}

We have provided a microscopical description for some of the configurations of
branes wrapping black hole horizons of \cite{DGMSWZ} in terms of gravitational
waves (gravitons) expanding to macroscopical configurations due to dielectric
effect. Interestingly, this dielectric effect is caused not by form flux but
only by gravitational effects.

The gravitational dielectric effect was studied in \cite{BGSW}. There it was argued that a gravitational dielectric effect for
D-branes should exist in non-trivial backgrounds for time-dependent
configurations. In our case we are considering gravitational waves, which
carry by construction momentum. This means that our configurations are in a
sense time dependent (although at the level of the effective action for the
gravitational waves we do not see it, since the momentum is added as $NT_0$ by
construction). We believe that the configurations presented in this paper are
nice examples of the ones proposed in \cite{BGSW}. In any case, we should point that, contrary to the configurations analyzed in \cite{BGSW}, ours are wrapping non-contractible cycles. This means that the pointlike solution is absent in this case. In any case, the construction presented here suggest that the philosophy in \cite{BGSW} that dielectric gravitational effect is still intact, in the sense that due to a non-trivial background, objects with non-zero velocity in a non-trivial background can polarize into a higher dimensional object, which is, for a large number of gravitons, effectively described as a macroscopic brane.

As mentioned in some points, the fact that our configurations wrap non-contractible cycles together with the fact that they sit in the lowest energy level in the internal torus indicates that the only possible instability is in the position in $AdS$. In general, simply by plotting the energies, it is easy to see that the extrema where our configurations sit are the absolute minima of the corresponding energy, and that any fluctuation would have more energy. In the type IIA case we can even explicitly see that the position of the minimum is related to the one half BPS condition, so at the end of the day it is SUSY what ensures the stability.

The main interest of these configurations is that they could be of some 
help
when understanding the black hole entropy, in very much of the same spirit of
what was proposed in \cite{GSY}. Since in the most general case the branes
carry momentum in the internal manifold, there is a coupling to the magnetic
potential, which becomes effectively a 2-form field strength on the 
spherical
part of the geometry. Then, the lowest energy branes wrapping the horizon
(which only contribute to net D0 or momentum charge) are in the
Lowest Landau Level of the construction. As shown in \cite{GSY}, in the
$AdS_2\times S^2$ black hole the degeneration of the
Lowest Landau Levels matches with the first order entropy formula for the
black hole. 

In this paper, we have
concentrated on a limited class of such branes wrapping the horizon, namely
the ones with zero momentum in the internal manifold (which are, as we have so
far seen, precisely the lowest energy ones), giving a successful
microscopical description of them. Although we have not presented it here, 
the extension to the most general case with momentum in the internal manifold
is straightforward, but hides in a sense the purely gravitational dielectric
effect responsible of these configurations. In any case, as we have shown, the
ansatz of static branes in the torus is fully consistent and implies that our
brane is indeed in the Lowest Landau Level.

Our main aim was to give a microscopical description of the branes in
\cite{DGMSWZ} in terms of gravitational waves. Therefore, in order to
have agreement with the macroscopical description and in very much of
the same spirit in \cite{M}, we considered a large $N$ limit. In any
case, nothing forbids us to consider the finite $N$ version, where our
description is fully non-commutative\footnote{Indeed since very
  recently it is technically possible to compute the symmetrized trace
  at finite $N$. See \cite{Costis}.}. Since this brane is wrapping the horizon, it seems that it becomes in some sense non-commutative. Although we cannot claim that our branes capture all the degrees of freedom of the horizon, it seems that since they wrap it and at finite $N$ they are non-commutative, the horizon must be effectively non-commutative. Focusing in the IIA case, where there is a better understanding, the conjecture in \cite{OSV} suggests that the black hole should be understood in terms of fixed magnetic charges and a weighted ensemble of electric charges, which led to the $AdS_2/CFT_1$ proposal in \cite{GSY}. In that reference, using this conjecture, they computed the leading order terms in the black hole entropy, finding an agreement with the macroscopic area law. It is argued that the main contribution to the entropy comes from the degeneracy of D0 branes expanding to the $S^2$. As we have so far seen, this branes indeed come with a big degeneracy since they are in Lowest Landau Levels (which are discrete). This is computed assuming a large number of electric charges, \textit{i.e.} zero branes. In such a situation the expanded D0 configuration becomes effectively (exactly as we have seen) a smooth D2 brane for which counting the degeneracy is easier. But indeed, as pointed in \cite{GSY}, there should be $\frac{1}{N}$ corrections to the entropy. In such a finite-$N$ situation, the expanded D0 system would be no-more an abelian D2, and effectively the horizon should become non-commutative.

We have considered the black string in M-theory, whose near horizon
geometry admits a stable brane wrapping the horizon, as pointed in
\cite{DGMSWZ}. By using the theory for coincident gravitons in \cite{JL2}, we
were able to give a microscopical description which, as expected, overlaps
with the macroscopical one when the number of gravitons is large. The black
string in 11 dimensions is directly related to the type IIA $AdS_2$. Upon
reduction of the 11d background we arrive to a geometry whose near horizon
limit is $AdS_2\times S^2\times T^6$. Since the direction in which we are
reducing is closely related to the one in which the 11 dimensional gravitons
which blow up to the $S^2$ wrapping the horizon in $AdS_3\times S^2\times T^6$
are moving,
we expect to have a similar phenomenon in IIA, but this time in terms of
dielectric D0 branes. Indeed this was analyzed in a
series of papers (\cite{OSV}, \cite{SSTY}, \cite{GSSY}) which led to a
proposal for the $AdS_2/CFT_1$ duality in \cite{GSY}. We have analyzed a
limited class for such branes, which are the ones which sit at the origin of
the $T^6$ (precisely due to this we could change the $T^6$ for a generic
$CY$). We have seen that the stability of the D0 brane system is not due to a
standard Myers effect, since there is no coupling to the 3-form. In turn,
there is a contribution from the CS action due to a monopole coupling of each
D0 to the 1-form. As we have argumented, when regarded from the 11-dimensional
point of view, this is inherited from the fact that in 11 dimensions the
configuration is stable only due to gravitational dielectric effect.

We have also considered the IIB $AdS_3\times S^3\times T^4$ background. Type IIB backgrounds have been considered in the
giant graviton literature in \cite{JLR1} and \cite{JLR2}. In those references
the microscopical theory for IIB gravitons was constructed. In this case an extra isometry is needed in the non-abelian version due to the
necesary T-duality in order to go from IIA to IIB. This isometry played an
important role, since due to its presence we were able to couple in the
$AdS_5\times S^5$ case the 4-form potential to the 1-dimensional worldvolume
of the gravitational waves (\cite{JLR1}). In \cite{JLR2}, the case of the
giant graviton in $AdS_3\times S^3\times T^4$ was analyzed. In that background
there was a 2-form RR potential, and therefore, in order to couple it, the presence of an
extra T-duality scalar was needed (see \cite{JLR2}). This T-duality extra
scalar played a key role in this giant graviton, since it represented the length of a non-commutative cylinder whose
sections are the physical giant gravitons (see \cite{JLR2} for details). In the case at
hand now we have a different situation, since there is no 2-form but 4-form
with entries in the torus. Therefore, once we take the ansatz that the
configuration is static in the torus, the RR potential plays no role. In any
case, since we are in type IIB, we have the extra isometry $l$ in the action. It is then natural to regard $S^3\sim S^1\rightarrow S^2$,
and identify the $S^1$ fiber with $l$. Therefore, although in $AdS_3\times
S^3\times T^4$, this construction is more similar to that of the giant in
$AdS_5\times S^5$ of \cite{JLR1}. In any case, we see that the extra isometry
required by T-duality is playing a key role, since it naturally allows us to
split the $S^1$ fiber from the $S^2$, which is the only pure non-abelian part
of the $S^3$.

It is instructive to analyze a configuration in which we had the brane wrapping the horizon with no velocity in
a cycle contained in the $AdS$. Then, the brane would tend
to collapse to the origin of the $AdS$, since the $AdS$ throat would pull it there. The presence
of momentum in the configuration compensates this collapse and renders the
brane stable far from the origin of the $AdS$. From this point of view, in the type IIA case, the coupling to
$C^{(1)}$ responsible from the stability of the brane can be seen as the
analogue to the ``centrifugal force'' on each gravitational wave in the 
11d
configuration. Therefore, in IIA, the configuration is stable not due to a
dielectric coupling, but due to a monopolar coupling which represents in a
sense the force on each D0 brane.

\vspace{1cm}
\noindent
\textbf{Acknowledgements}\\

\vspace{-.3cm}
It is a pleasure to thank Bert Janssen, Yolanda Lozano, Jan de Boer, Sergio Monta\~{n}ez,
Marcos Su\'arez and Carlos Hoyos for fruitful conversations. 
This work has been partially supported by CICYT grant BFM2003-00313
(Spain). The author was supported in part by a F.P.U. Fellowship from
M.E.C. (Spain).

\end{document}